# Synchronized Multi-Load Balancer with Fault Tolerance in Cloud

Sreelekshmi S[1], K R Remesh Babu[2]

[1] Department of Information Technology,
Government Engineering College Painavu, Idukki, Kerala, India
*sreelekshmisuresh94@gmail.com*

[2] Department of Information Technology,
Government Engineering College Painavu, Idukki, Kerala, India
*remeshbabu@yahoo.com*

*Abstract*: Cloud computing offers on-demand access to a large pool of shared resources at lower cost. The advantage of cloud resources is that it can be easily provisioned, configurable, and managed with minimal management efforts by the users. Proper load balancing is an important task in maintaining fault tolerance and Quality of Service (QoS). In the cloud, a load balancer accepts incoming user requests, application specific traffic and distributes this workload across multiple backend processes using various methods. In a single load balancer system; if the load balancer is down none of the user tasks can't be processed, even when the servers are ready to process the tasks. In order to overcome this single point of failure, this paper proposes a model that will avoid the single point of failure by using multiple load balancers. In this method, service of one load balancer can be borrowed or shared among other load balancers when any correction is needed in the estimation of the load. This will improves fault tolerance of the cloud eco system and assist in cluster capacity management.

## I. Introduction

With the emergence of cloud computing more and more business organizations moving towards cloud computing platform due to its attractive features like low cost, easily configurable, and virtually unlimited resource pool with on-demand provisioning. The performance of the cloud eco system enhances, if the scheduling of resources is properly done. Resource scheduling with load balancing is one of the best methods for improving the cloud performance. The researchers are proposed several methods for optimal scheduling of resources in the cloud.

Resource optimization [27] is the process of efficient utilization of the available resources. It achieves desired results within a time span and budget with minimum usage of the resources. It has the following benefits:
- Increased revenue: The resource management solutions ensuring the most valuable resources are to be used in a maximum effect.

- Boost efficiency: The optimization leads to more efficient utilization of the resources.
- High quality results: Optimization can reduce number of errors and achieving better results.
- Security: The proper optimizations allow a secure environment. That is the optimized results can reduce the risk of data processing

Even though the features of clouds are attractive and there is in need of a fault tolerant mechanism to undisrupted performance of cloud services. Load balancing mechanism [28] can improve the performance by efficient distribution of workloads across multiple computing resources such as computers, network links or disk drives. Tasks received by a load balancer can be distributed to any cluster members. Numerous techniques are available for the distribution of workload across processors and the optimal scheduling leads to the optimal result. The factors considered for these optimization techniques are different. Some of the optimization condition for the task distribution includes the minimum response time, energy consumption and maximum profit benefits. Load balanced cluster is an abstraction for a set of identical processors, that host same set of services. A simple cluster with a load balancer and respective cluster members are shown in Fig. 1. Here n servers are managed by a single load balancer. When this single load balancer fails or down, the entire system functioning collapses due to the non-availability of the load balancer cum dispatcher. This will cause financial as well as loss of credibility of the cloud provider.

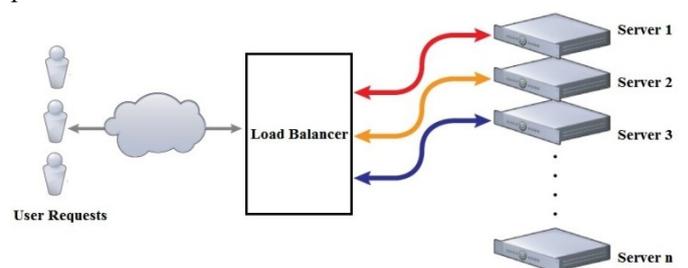

**Figure 1.** Simple cloud load balancing

Load balancing is one of the overriding issues in cloud computing due to the dynamic nature of the cloud. As in a distributed environment, load balancing mechanism in cloud distributes the dynamic workload evenly across all the nodes in the cloud to avoid a situation where some nodes are heavily loaded while others are idle or doing little work. It helps to attain increased satisfaction to the customers and high resource utilization that consequently improving the overall performance and profit of the provider. When the load balancer is down, the entire process will be crashed even when the processor is ready to process the task. It causes a single point of failure for the entire system.

In the mission critical application of single point failures are to be avoided. If we adopt multiple load balancers, it will increase the fault tolerance of the system. This paper proposes a modified fault tolerant system with optimized scheduling that can improve the existing mechanism in load balancing with capacity estimation [1]. Sliding window based self-learning and adaptive load balancer (SSAL) [7] is an observation based load balancer that can produce optimal throughput in both stable and unstable environments. SSAL monitored the performance of the cluster members in every feedback interval and is trying to overcome the problems due to single point failure. Also it is used to make corrections in the load distribution model.

The main contributions of this paper are (1) Single point failure of the system can be recovered by the usage of multiple systems in parallel (2) Sharing of load balancing information among all other load balancers and (3) An analysis to find out cluster capacity needed for the better performance of the system.

This proposed work is organized as follows. Similar works are already proposed by the researchers are reviewed in section 2. Problem identification, detailed design and explanation of the proposed method are described in section 3 and section 4 respectively. In section 5 covers the performance analysis and finally the paper concludes in section 6.

## II. Related Works

Load balancing, is one of the important and difficult areas of cloud computing. The load is unpredictable in cloud computing and it can be varied, depending on the demand for a particular service. For ensuring better performance and QoS, the load balancing mechanisms have more important role. There are several papers are available related to this issue. Fault tolerance is also a significant issue in parallel applications. The paper [2] gives an idea about fault tolerant parallelization with task pool pattern in global load balancing. Also describe a fault tolerant mechanism in paper [3]. Here uniformly dispense the workload across the nodes and eliminates the faults from the network. It contains a frame work for tolerating simultaneous failures. For handling the dynamic load among the virtual machine, an efficient load balancing of resources is necessary. A fault tolerant load balancing techniques based on a graph structure is illustrated in article [4]. The model can improve the utilization of available resources in the environment along with fault tolerance. Service level Agreement (SLA) is an agreement between the customer and service provider. It develops a prevention method for SLA violation to avoid costly penalties' [5]. In grid and cloud computing the role of load balancer is important to deal with potential problems, such as high level of scalability and heterogeneity of computing resources [6]. Here present a generic load balancing scheme, which separates the allocating and migrating process while preserving a guaranteed level of service. The work in paper [8] provides different load balancing and job migration techniques for scheduling tasks. In the virtualized scenario, task scheduling can also be performed using preemption and non-preemption based on the user requirement.

Task allocation and scheduling on a set of virtual machines is one of the important difficulties in cloud computing. It can be overcome with heuristic algorithms, which includes Genetic Algorithm (GA), Particle Swarm Optimization (PSO), Ant Colony Optimization (ACO) etc. Task allocation with an efficient greedy algorithm and genetic algorithm with the help of cross over and mutations is described in [9].Virtual Machine scheduling in cloud environment proposed in [29] is a model of VM load balancing based on task execution time span. Multi objective method for the optimal work load distribution using particle swarm optimization [10] can minimize the response time and cost of the incoming request and maximize the profit of the broker. The resource allocation performed with the help of genetic algorithm is presented in [11]. Here the optimization of incoming VM request by minimizing the response time (RT) and Cost of VM instances to maximize the profit of the broker. The paper [12] provides an optimal scheduling with energy efficient method without crossing any uncomfortable delay to the customer. Markov decision process [13] can also be used for the optimal scheduling of energy storage devices in power distribution network with minimizing cost of energy.

When the central part of the system is down, the overall performance of the system is degraded. It is known as the single point of failure. A method for overcome this single point of failure using heart beat algorithm is illustrated in [14]. There are different ways to balance the load optimally. Paper [15] provides a survey related to the optimization techniques based on evolutionary and swarm based algorithms. An algorithm called Multiple Agent-based Load Balancing Algorithm (MA) in which shifting of the workload is carried out in IaaS cloud to achieve well dynamic load balancing across virtual machines for maximizing the resource utilization [7]. A novel algorithm for sharing distributed file systems is proposed in paper [16]. Here, nodes are simultaneously serves computing and storage functions. A File is partitioned into a number of chunks and is allocated to distinct nodes so that tasks can be performed in parallel over the nodes. The paper [17] discusses and compares load balancing algorithms to provide an overview of the latest approaches in this field. Paper [18] proposes a load balancer framework, which is aware of multiple quality of service, in large scale distributed computing system. The review in paper [19] aims to provide a structured and comprehensive overview of the research on load balancing algorithms in cloud computing. The vital part of this paper is the comparison of different algorithms considering the characteristics like fairness, throughput, fault tolerance, overhead, performance, and response time and resource utilization. Based on the load status, the system can dynamically shift the load from the heavily loaded controller to the lightly loaded ones [31]. An open flow based dynamic traffic scheduling takes the advantages of Software Define Network (SDN) central controllers [32].

The load balancing can be performed by different algorithms. These algorithms are classified into the static, dynamic, bio or nature inspired, and game theory based algorithms. The static algorithm includes random algorithm, round robin algorithm, min-min, min-max algorithm and weighted round robin algorithms. In the methodical analysis of various balancer conditions on public cloud division, Ant colony and Honeybee behavior is best for the balancing of load under normal balancer condition [20]. In idle balancer condition round-robin is being applied which appears suitable for that condition. In addition for huge and complex corporate area, it focuses on the strategy of divisions based on region to simplify the load balancing. The relation between probabilistic routing and weighted round robin load balancing policies is explored in [21]. Cloud computing issues like resource provisioning, load imbalance and performance improvement can be solved using bio-inspired algorithms. Paper [22] gives a detailed review of the bio-inspired algorithms proposed in cloud computing. Genetic algorithm is a search algorithm based on the principles of evolution and natural genetics'. The work [23] proposes a GA based load balancing strategy for cloud computing.

In order to improve resource utilization and profit, more number of VMs are allocated to a particular server, the performance delay will create interference [33] and that will affect overall QoS. The article [24] gives an idea about QoS of multi-instance applications in the Clouds. This approach is based on limiting the number of requests at a given time that can be effectively sent and stored in queues of virtual machines through a load balancer equipped with a queue for incoming user request. The paper [25] proposes a QoS aware load balancing scheme in congested extended service set environment. A QoS-aware replica placement for data intensive applications is presented in paper [26]. It addresses the QoS aware replica placement problem in the data grid, and proposes a dynamic programming based replica placement algorithm.

## III. Problem Identification

One of the main features of the cloud is that, on-demand computing at any time at low cost with ensured QoS. In the cloud, there is no explicit knowledge for the customer about where the task is being executed and in which server. Cloud providers are trying to offer fault tolerant service to their customers. But single point failures are one of the barriers for fault tolerant continuous service. Since the load balancer is responsible for distributing the tasks received from the end users to the optimal processors by considering the minimal response time, energy consumption and maximum profit earned. The processing of tasks will be halted when the load balancer is down, even when the processor ready to execute it.

This may be due to the hardware failures like, server crashes, network down, power failures or disk crashes. Software failures like directory proxy server crash and database corruption will also result in single point failure. So to address these failures, a suitable cooperative mechanism is needed for fault tolerant cloud service.

## IV. System Design

The proposed system contains a number of schedulers (load balancers) and each scheduler can able to balance the task across multiple processors. These schedulers interacts each other to communicate the information they gathered about the running task status and their tasks in the input queue. They are also able to distribute the tasks to other processors in the data center based on the known capability of each processor. After the execution of each processor, it can generate a feedback based on the current capacity of each server. The capacity calculation is done in fixed time interval based on number of tasks processed by the processor and the tasks pending in the queue.

This frequent monitoring and cooperative load balancers ensure the QoS to the end users. Also in cooperative load balancing, none of the load balancers are overloaded due to the sharing of information about tasks already completed, being executed and waiting in the queues. The architecture of the proposed method is shown in figure 2. The detailed explanation is given in the next sub sections.

### A. Task Handling

A set of tasks with distinct specifications from the end users are to be handled by a task handler, in which identical tasks are eliminated and the remaining are stored in it. Hence it can reduce the overhead of the entire process by removing duplicate tasks. An SLA checking based on the cost and time constraints are to be performed in this level. Here considering the user specified cost of the incoming task with the price of the service provider. If the deviation is greater, then the corresponding requests are accepted otherwise there exists an SLA violation of the task.

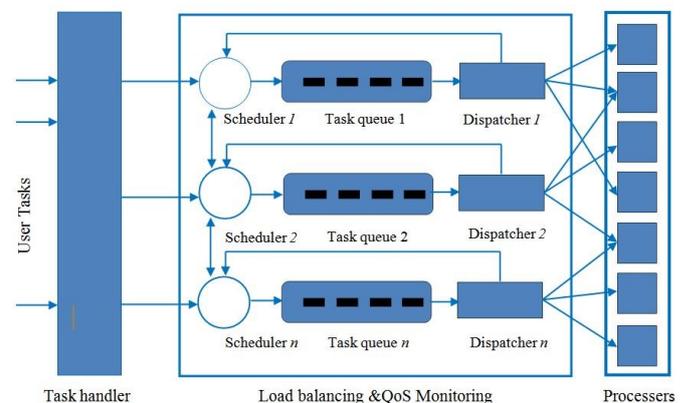

**Figure 2.** System model

The time based mechanism which considers tasks' arrival time and response time. If the difference is small, such tasks are to be accepted others are rejected or SLA violation take place if accepted. Then the tasks are distributed among different load balancers on the basis of round robin scheme. Once the scheduler is down, the tasks stored in the queue are transferred back to the request handler. That is the task handler stays active until the completion of the processes at each scheduler.

### B. Load Balancing and Capacity Calculation

A set of task is given to each scheduler (load balancer), which stores them in an output queue. There exists a dispatcher for distributing the requests to different processors based on individual processing capacity. It is computed on the basis of three constraints. When a client submits a tasks to the service provider through an intermediate cloud broker, the client want to complete the job in a short period of time. Therefore response time can be considered as one of the objective function.

$$\text{Response Time} = \text{Transmission Time} + \text{Processing Time}$$
$$= (T_s/bw) + (T_s/P_s) \quad\quad\quad (1)$$

When the user submits a tasks to the service provider, the cloud broker find the best solution for the user satisfaction. During the process, the broker expected to obtain a certain profit. Therefore, maximizing profit of the broker can considered as the second objective.

$$\begin{aligned}\text{Profit} &= \text{Processing cost of PM} - \text{Cost of user task} \\ &= P_c - T_c \\ &= (P_t * P_{pm}) - T_c \\ &= (T_s/P_s) * P_{pm} - T_c \quad\quad\quad (2)\end{aligned}$$

For the processing of tasks from the user, the service provider needs an energy usage. $E_j$ is the energy consumption of service provider $j$ to execute a task. Minimum consumption of energy can be considered as the third objectives

- $T_s$   - Task size
- bw   - Bandwidth of the processor
- $P_c$   - Processing cost
- $T_c$   - Task cost
- $P_t$   - Processing time of PM
- $P_{pm}$   - Price of PM

From the available information it can find a processor with minimum response time, minimum energy consumption and maximum profit that can be earned for processing tasks on a service provider. This can be computed on the basis of a ranking strategy. Ranking procedure is considering the response time, energy consumption and profit earned during the processing of each task in each processor. Tasks can be assigned on the basis of available resources in server and considering the requirement of incoming tasks. Then it finds the optimal processor for each task by considering the optimization condition. An example for ranking strategy is shown in Table 1 and 2. The capability may be varied under special circumstances like the processor being down or crashed or some heavy load is being executed on the servers. The resource capability correction is handled by a single scheduler (known as the coordinator) selected from the set of schedulers, based on a centralized method. The central coordinator can make corrections based on the observation reported by individual schedulers. The coordinator is selected in accordance with the algorithm given in figure 4. While tasks are being executed in different processors, the dispatcher makes a feedback to the schedulers regarding the new capability of processors.

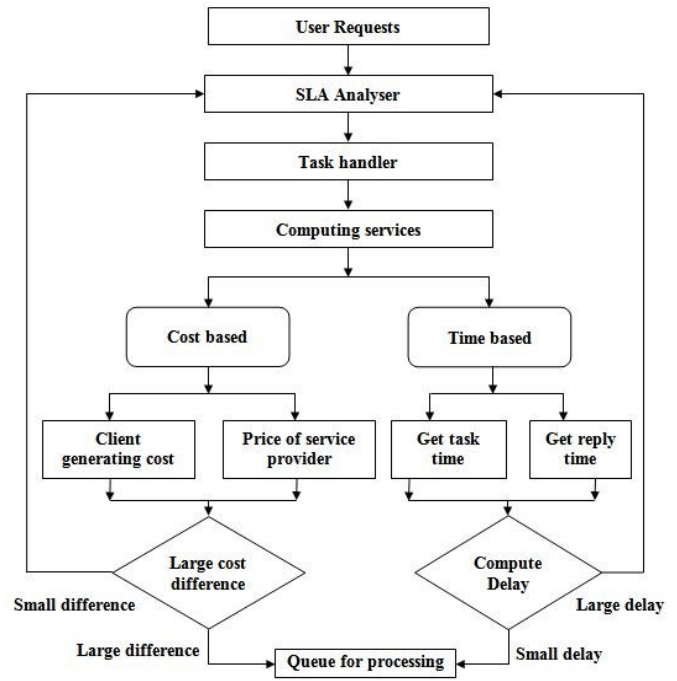

**Figure 3.** SLA checking

|  | $P_1$ | $P_2$ | $P_3$ |
|---|---|---|---|
| **Task 1** | (8, 100, 100) | (12, 15, 50) | (14, 10, 30) |
| **Task 2** | (9, 20, 40) | (19, 10, 90) | (21, 50, 50) |
| **Task 3** | (9, 10, 10) | (21, 50, 50) | (11, 40, 70) |

*Table 1.* Before ranking strategy.

|  | $P_1$ | $P_2$ | $P_3$ |
|---|---|---|---|
| **Task 1** | (8, 100, 100) | (12, 15, 50) | (14, 10, 30) |
| **Task 2** | (9, 20, 40) | (19, 10, 90) | (21, 50, 50) |
| **Task 3** | (9, 10, 10) | (21, 50, 50) | (11, 40, 70) |

*Table 2.* After ranking strategy.

Here in Table 1 and 2 each entry (a, b, c) is (Response time, Energy consumption, Profit) and $P_1$, $P_2$, $P_3$ represents Processors. From this the optimized best result is: Task 1 – $P_1$, Task 2 – $P_2$, Task 3 – $P_1$.

1. Begin
2. Multicast coordinator selection information and the time is noticed.
3. If no message is received from other schedulers. current one becomes the coordinator
4. If message received, the reporting time is noticed and the scheduler with greater responding is selected as the coordinator.
5. If more than one of them has the same responding time then scheduler with the highest capacity is selected as the coordinator
6. The selected coordinator details multicasts to all others.
7. Return

**Figure 4.** Coordinator algorithm

The feedback contains information about the number of tasks processed by the processer and those are pending in the output queues of the respective processor. Initially, one scheduler acts as the coordinator. The coordinator process the information based on the algorithm given in figure 5. It also calculates the capability of each processor. The coordinator now multicasts the capability information obtained to every other scheduler in the data center. In the next stage, all the schedulers work in parallel using this capability information. Each of the individual schedulers obtains the capability information from the processor as a feedback. These schedulers pass the obtained information to the coordinator, for performing the necessary corrections. This will be done by the generation of the capability information of individual processor in an updated manner, using the possible combinations of capabilities provided by schedulers in different instances of time. Based on the newly available capacity information, it can distribute the tasks among processors. In this method, the coordinator is assumed to be down, when any of the schedulers do not obtain the information in three consecutive multicasts.

1. Begin
2. Select coordinator
3. Each scheduler ($S_i$) monitors the number of Tasks processed ($X_{ijt}$) and the number of tasks in the queue ($Y_{ijt}$) for the processor, for every feedback interval ($t$)
4. After feedback interval send the information to the coordinator
5. Coordinator collect the observation reported by each scheduler
6. Coordinator Calculate average no of tasks processed by each processor($P_j$) at interval($t$) $AP_{jt} = \sum_{i=1}^{n} (AX_{ijt})$
7. Calculate the average no of tasks pending in the queue processed by the processor ($P_j$) as $PR_{jt} = \sum_{i=1}^{n} (AY_{ijt})$
8. Estimated request for processor $P_j$ is $ER_{jt} = AP_{jt}$
9. Estimated capability of the processor $P_j$, $EC_{jt} = (ER_{jt}/\max(1, PR_{jt}))$
10. Relative capability of the processor $RC_{jt} = (ER_{jt} / \sum_{i=1}^{n} (EC_{jt}))$
11. Total tasks to be issued in the next feedback interval $T_t = \sum_{j=1}^{n} ER$
12. Total tasks issued for the next feedback interval by the scheduler is $T_{it} = (T_t)/n$
13. Send adjust load distribution message to all schedulers
14. Return

**Figure 5.** Capacity estimation algorithm

Where $X_{ijt}$ is the number of tasks handled by the scheduler $S_i$ to processor $P_j$ in feedback interval $t$ and $Y_{ijt}$ is the number of tasks pending in the queue of processor $P_j$ at scheduler $S_i$ in the interval $t$ and $n$ is the total number of schedulers.

For better performance of the system, each scheduler can monitor the throughput of the incoming request and make a comparison with a standard value. Based on the information from the schedulers, the coordinator can make a correction in the cluster capacity.

Standard value (β) is generated based on the total cluster capacity (CC) and the total number of tasks (T) in the task handler at time $t$. It is calculated using the equation (3).

$$\beta = (CC/T_t) \qquad (3)$$

Each scheduler monitors the throughput value for each task and compares them with the standard value. Also, schedulers calculate the deviations from these values. If it is above β then considers it as a Success Variation (SV), if it is the below the limit then consider it as a Failure Variation (FV) for each task. Equation (4) and (5) is used to calculate SV and FV of a $i^{th}$ scheduler for a request $j$.

$$SV_{ij} = (\beta - \text{measured value})/\beta \qquad (4)$$

$$FV_{ij} = (\text{measured value} - \beta)/\beta \qquad (5)$$

Over time capacity $OC_t$ at a particular time $t$ is the sum of success variations and it is represented by equation (6). Under capacity at time $t$ ($UC_t$) is the sum of failure variation over the limit. It is calculated using the equation (7).

$$OC_t = \sum_{i=1}^{n} SV_{ij} \qquad (6)$$

$$UC_t = \sum_{i=1}^{n} FV_{ij} \qquad (7)$$

Capacity Deviation (CD) is the difference between over and under capacity. Then the increase in cluster capacity is determined by the equation (8).

$$\text{Increase the cluster capacity} = (CC/TR_t) * CD \qquad (8)$$

Where $TR_t$ is the total request to a scheduler. If the under capacity is greater than the over capacity, the cluster capacity can be reduced using the equation (9).

$$\text{Capacity Reduction} = (CD * (CC/(TR_t + CD))) \qquad (9)$$

## V. Experimental Setup and Results

The proposed method is simulated using CloudSim 4.01 with three schedulers. In the initial stage, one scheduler is used for distributing the entire request to the servers. After this initial step, schedulers calculate the capacity of every server using the capacity calculation algorithm. Based on the newly measured capacity, all the schedulers can distribute the load across servers.

From the capacity deviation analysis shown in figure 6, the deviation is gradually increased when the number of the task is increasing. When the number of tasks is 20 the deviation is 33.33%. so the system needs 33.33% or additional resources for effective load balancing. Similarly 60%, 66.67% and 57.20% when the number of the task are 50, 60 and 70 respectively. Also note that in the initial stage, there are no deviations in the load, due to fewer users are present.

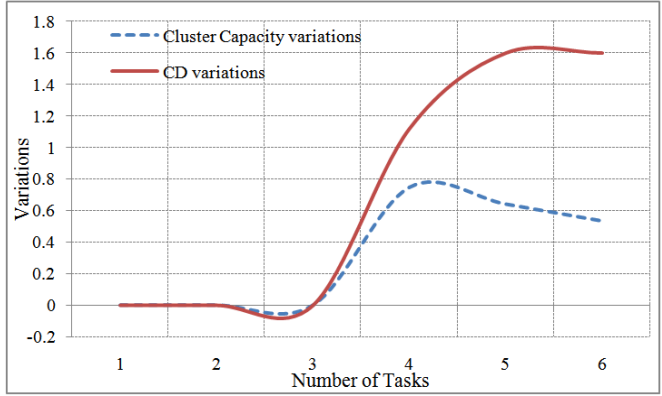

**Figure 6.** Variations graph for 2 VMs

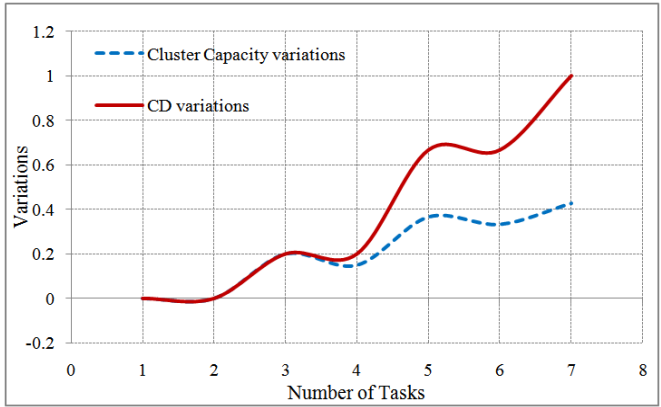

**Figure 7.** Variations graph for 3 VMs

From the capacity deviation analysis shown in the figure 7, the deviation is gradually increasing with the increase in number of tasks. When number of tasks is 40 then there occured a deviation of 25.00%. The system require 45% or additional resources for the scheduling process. In this way when the number of task are 60 and 70 the corresponding fluctuations are 50% and 47.16% respectively. Also note that the execution of three VMs leads to a 7% of decrement of cluster capacitywhen compare to the usage of two VMs.

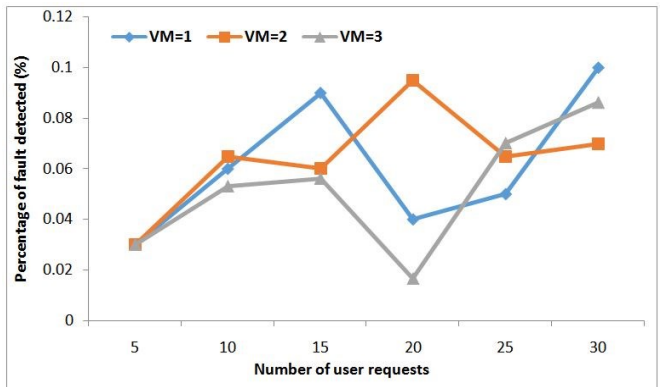

**Figure 8.** Number of fault of occurred

Figure 8 shows the percentage of fault occurred for different number of VMs with number of user requests. From the figure it is observed that number of faults occurred is less than 0.1% in all the cases. This shows the effectiveness of the mechanism, i.e., the proposed cluster variation mechanisms gives nearly 99.9% fault tolerant execution of user requests.

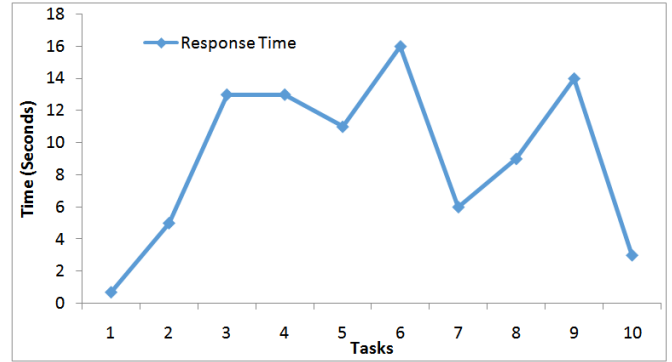

**Figure 9.** Response Time variation

Similarly response time variations are measured for different number of user requests for a single VM is shown in the figure 9. As the number of requests increases there is no significant variation in the response time.

The cost benefit analysis for the proposed method is given in figure 10. The fault tolerant execution is cost effective for the provider when the providers have minimum number of active users. Our experiment shows that when the number of users are too high or very low, the provider is not in a better position. This is due to two conditions. (1) at low load, the provider have to run more number of physical servers to maintain QoS and (2) at high low load, the penalty is high due to possibility of SLA breaches. This can be avoided using suitable migration and auto scaling techniques. So in future, a fault tolerant system with suitable auto scaling mechanism needed to accommodate more number of users.

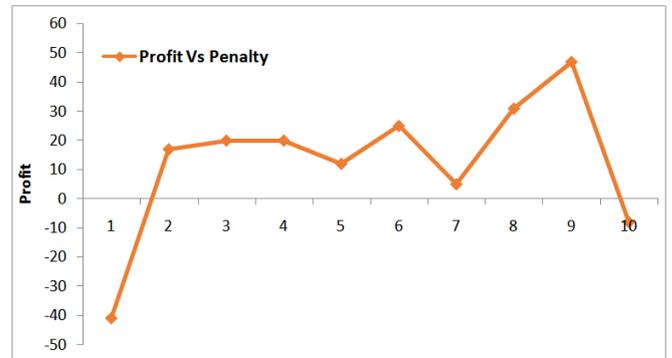

**Figure 10.** Cost Benefit analysis

## VI. Conclusion

The load balancer receives the request and distributes to servers which has minimum Response time, minimum energy consumption and maximum profit to process them efficiently. When the load balancer fails, the user requests will not reach the servers and results the single point of failure for the overall system. Here propose a "Synchronized Multi-Load Balancer with Fault Tolerance in Cloud" that extends the single load balancer to make it more fault tolerant. The estimated cluster information shared among different user groups to collaborate multiple schedulers for fault tolerance. The scheduler also provides additional functionality to set and monitor the performance standards and find the cluster capacity changes needed to meet the standard value. In future it can be extended for energy aware scheduling.

## Author Biographies

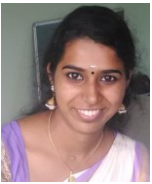

**Sreelekshmi S**, was born in Kerala, India in 1994. She received the B Tech Degree in Computer Science and engineering from the Cochin University Of Science And Technology (CUSAT), Kerala, India in 2016. She is currently perusing Masters Degree (M Tech) in Network Engineering from APJ Abdul Kalam Technological University, Kerala, India. Her research interest includes, Distributed and cloud Computing, Big data analytics, and Internet of Things.

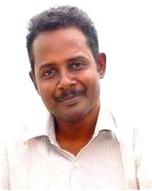

**K R Remesh Babu** received his BSc. degree in Mathematics from Mahatma Gandhi University, Kottayam, India and B.Tech in Information Technology from Cochin University of Science & Technology (CUSAT), Kochi, India. He holds ME in Computer Science from PSG Tech Coimbatore, India. He is currently pursuing the Ph.D. degree at CUSAT. He is an assistant professor in department of Information Technology, Government Engineering College Idukki, India. He has published more than 35 research papers in International Conferences and Journals. His research interests includes Distributed and Cloud Computing, Internet of Things, Wireless Sensor Networks, and Big Data Analytics..